# A Method for Increasing the Resolution of Optical Detectors

S. Khademi[1], A. Darudi[1] and M. Shahsavari[2]
[1]Department of Physics, University of Zanjan, University Bulverde, P.O. Box 313, Zanjan, Iran.
E-mail: skhademi@mail.znu.ac.ir
[2] Geophysics Institute, University of Tehran, Amir-Abad Street, Tehran, Iran.

**Abstract:** In addition to the optical aberrations, the magnitude of the optical cells of detectors is one of the most important parameters, which restricts the resolution of detectors. Adaptive optic and the methods of reducing the aberrations are often used to increase the resolution of optical systems. In the best situation the image structures, which are larger than the magnitude of the cells, are detected and the finer structures are removed. In this paper, a new method for increasing the resolution of images is presented. In this method, the cells will scan the images by a piezoelectric crystal. The piezoelectric crystal moves the cells in $n$ identical steps. In each step, cells move $\frac{1}{n}$ of the cell magnitude and the data are recorded. Finally, data are analyzed and the structure of images is reconstructed. In this method, the structure of images is $n$ times finer than the cells magnitude

## 1. INTRODUCTION

In addition to the optical aberrations and diffraction, the pixel size of the optical detectors is one of the most important parameters which restrict the resolution of the optical systems [1]. In many situations the image structures are smaller than the pixel size. Thus, the structures of image which are finer than the dimension of pixel are not detected and hence the maximum limit of resolution is defined by the pixel size of detector. To have a better optical resolution one may choose a detector which has more pixels in the unit length but it is more expensive.

Recently the sub pixel resolution methods are developed by authors [2-4]. In this paper we investigate the one dimensional case and introduce an algorithm and simulation for increasing the resolution of the optical detectors (IROD) . The pixels virtually divided by $n$ parts. The detector is moved step by step, by a translator. After each step, the intensity of the pixels is recorded and after all there are $n$ strings of data. These data will be analyzed to obtain an image which the resolution is enhanced.

## 2. ALGORITM FOR INCREASING THE RESOLUTION OF THE OPTICAL DETECTORS

Suppose that the pixel size of an array detector is $L$ and the number of pixels is $N$. The detector is moved in discrete steps by a translator. The limit of movement is $L$ and amount of steps are $n$. The intensity of the pixel number $j$, which is denoted by $I'(j, m')$, will be recorded in each step. After a complete movement one has $n$ strings of data $I'(j, m')$. In this method each pixel is virtually divided by $n$ sub pixel or virtual pixel. The aim of this article is to reconstruct the intensity of virtual pixels which is denoted by $I(m)$, where $1 \leq m \leq nN$. In the following, we present an algorithm to calculate the intensity of the virtual pixels of image, $I(m)$ to increase the resolution from the recorded intensities $I'(j, m')$.

**Stage1:** In the first stage the intensities of the pixels, which are due to an image, are recorded in the string $I'(j, m' = 1)$. Then the translator will move the pixels by one step and the pixels will record the intensities again. One may continue this procedure $n$ times to scan the total length of a pixel.
For $j = 1$ one has:

$$I'(j = 1, m' = 1) = I(m = 1), \tag{1}$$

$$I'(j=1, m'=2) = \sum_{i=1}^{2} I(i),$$
$$\vdots$$
$$I'(j=1, m'=n) = \sum_{i=1}^{n} I(i), \tag{2}$$

and for $j=2$:
$$I'(j=2, m'=1) = \sum_{i=2}^{n+1} I(i),$$
$$\vdots$$
$$I'(j=2, m'=k) = \sum_{i=k+1}^{n+k} I(i),$$
$$\vdots$$
$$I'(j=2, m'=n) = \sum_{i=n+1}^{2n} I(i). \tag{3}$$

Follow the above procedure to obtain all data of $I'(j,m')$:
$$I'(j,m') = \begin{cases} \sum_{i=1}^{m'} I(i), & for \quad j=1, \\ \sum_{i=(j-2)n+m+1}^{(j-1)n+m'} I(i) & for \quad j>1. \end{cases} \tag{4}$$

Finally one will obtain $I'(j,m')$. For example $I'(4,6)$ is the intensity of the 4$^{th}$ pixel in the 6$^{th}$ step.

**Stage2:** The previous stage described the data recording while the next ones are concerned with the reconstruction of the images. After recording one may calculate the intensity of the $n$ virtual pixels of the first pixel. These values are the initial conditions for the calculation of the next virtual cells. At each stage one obtains the intensity of the virtual pixels $I(m)$ which contains the finer structure of the image in comparison to the ordinary one. To obtain the intensity of m$^{th}$ virtual pixel, one should know where the information of this virtual cell in $I'(j,m')$ is.

Therefore one will find the $j$ and $m'$ corresponding to each $m$. For the m$^{th}$ virtual pixel one has
$$\frac{m}{n} = [\frac{m}{n}] + \frac{m'}{n}, \tag{5}$$

where $[\frac{m}{n}]$ and $m'$ are the quotient and the rest of the division $\frac{m}{n}$, respectively. Thus for the next application in the algorithm of IROD method one needs $j$ and $m'$ in terms of $m$ as

$$j = \left[\frac{m}{n}\right] + 1, \tag{6}$$

and
$$m' = \begin{cases} m - n(j-1) & for \quad m' \neq 0, \\ n & for \quad m' = 0. \end{cases} \tag{7}$$

**Stage 3:** To construct the intensity of the virtual pixels one should follow the following straightforward procedure

$$I(m=1) = I'(j=1, m'=m=1), \quad for \quad m=1, \tag{8}$$

and
$$I(m) = I'(j=1, m'=m) - \sum_{i=1}^{m-1} I(i), \tag{9}$$

for $1<m<n$. Then for bigger values of $m$ one has
$$I(m) = I'(j=[\frac{m}{n}]+1, m') - \sum_{i=m-n+1}^{m-1} I(i) \quad for \quad m \geq n. \tag{10}$$

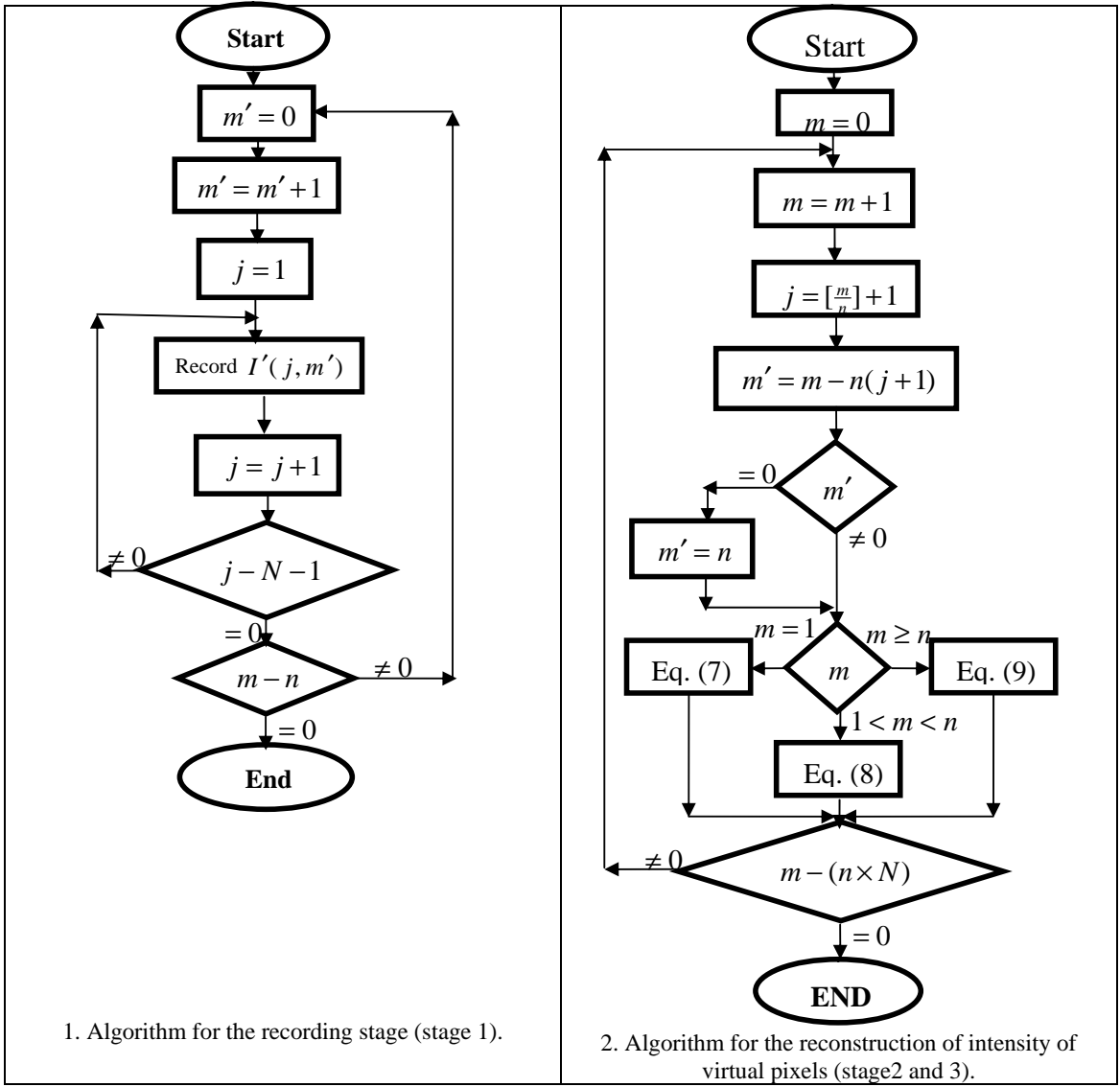

1. Algorithm for the recording stage (stage 1).

2. Algorithm for the reconstruction of intensity of virtual pixels (stage2 and 3).

## 3. SIMULATION RESULT

A computer simulation is used to illustrate the ability of the IROD algorithm. A reference image is generated by a MATLAB program and a graph of its intensity in the horizontal direction are shown in Fig.(1a) and (1b) respectively. The reference image has a wide range of frequencies. Figure (2a) is a simulation of the recorded intensity of a low resolution CCD. Figure (2b) is a graph of the Fig. (2a), in horizontal direction. Each pixel of the CCD is equivalent to e.g. 9 pixel of the reference image. To reconstruct a high resolution image, one may generate 9 images, $I'(j, m')$, by the virtual movement of the CCD where the size of each step is, e.g. $\frac{1}{9}$ pixel.

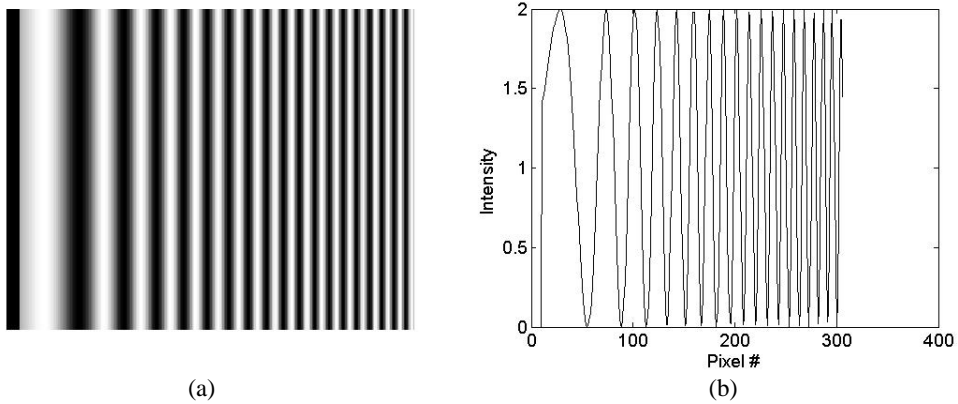

Figure 1: a) The reference image, b) A graph of the intensity in the horizontal direction.

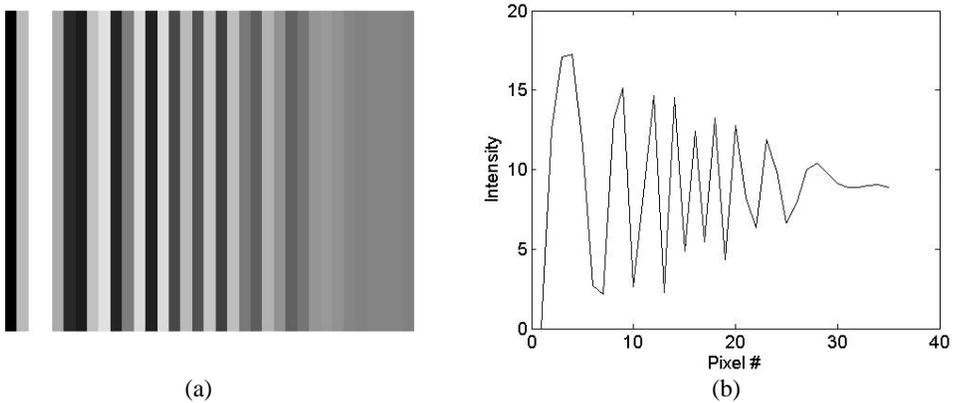

Figure 2: a) A simulation of the recorded intensity of a low resolution CCD. b) A graph of the intensity in horizontal direction.

Figure (3a) is a graph which is reconstructed by the IROD algorithm. The difference between the reference and the reconstructed graphs is shown is fig. (3b). As seen from fig.3b the error is too low.

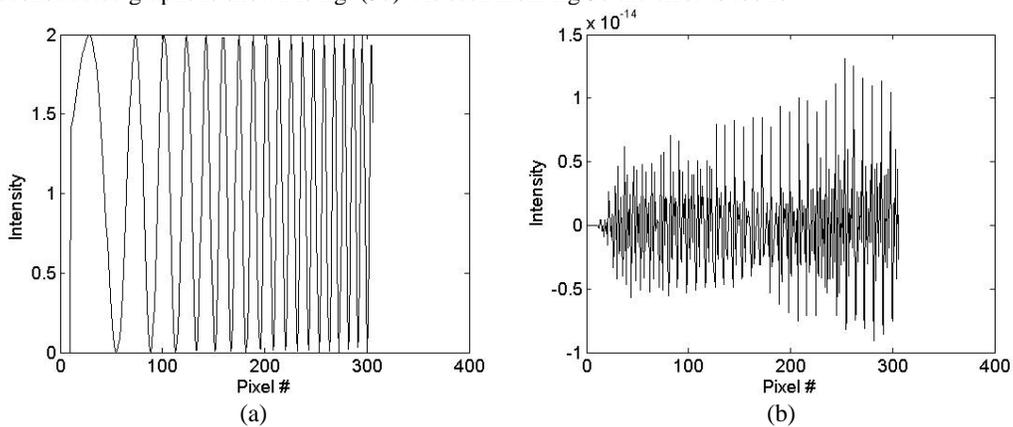

Figure 3: a) Reconstructed graph by IROD algorithm, b) Difference between the reference and the reconstructed graphs.

Actually, the movement error, nonlinear response of detector and the noises, which are not included in our simulation, may affect the results.

## 3. CONCLUSION

In this paper the theory of IROD method and simulation are introduced for an array detector and the algorithms for the recording images and reconstruction of the fine structure of the image is presented. The IROD method increases the resolution of the ordinary detectors $n$ times, where $n$ is the number of the sub pixels. The authors believe that, this method will considerably decrease the cost of making accurate detectors. The experimental work and development of this method for two dimensional case is investigated by authors and will be reported later.